\documentclass[a4paper,UKenglish,cleveref, autoref, thm-restate]{lipics-v2021}

\pdfoutput=1 
\hideLIPIcs  

\usepackage[textsize=footnotesize]{todonotes} 
\usepackage{xargs} 
\usepackage{xspace} 
\usepackage[linesnumbered,ruled,vlined]{algorithm2e} 
\usepackage{url}
\usepackage[bibliography=common]{apxproof} 
\usepackage{arydshln} 
\newcommand{\br}[1]{\boldsymbol{\color{red} #1}}
\newcommand{\E}{\mathcal{E}}
\renewcommand{\O}[1]{\mathcal{O}\left(#1\right)}

\newcommand{\ie}{{\it i.e.}~}
\newcommand{\households}{{\it Households}\xspace}
\newcommand{\highschool}{{\it Highschool}\xspace}
\newcommand{\infectious}{{\it Infectious}\xspace}
\newcommand{\foursquare}{{\it Foursquare}\xspace}
\newcommand{\wikipedia}{{\it Wikipedia}\xspace}
\newtheoremrep{theoremApx}[theorem]{Theorem}
\newtheoremrep{corollaryApx}[corollary]{Corollary}
\newcommand{\Output}[1]{{\bf output} #1}
\SetKwProg{Fn}{Function}{:}{}
\SetKwFunction{Union}{Union}
\SetKwFunction{Find}{Find}
\SetKwFunction{UF}{UF}
\SetKwFunction{MakeSet}{MakeSet}
\SetKwFunction{timeUF}{TimeUF}

\title{LSCPM: communities in massive real-world Link Streams by Clique Percolation Method}

\author{Alexis Baudin}{Sorbonne Université, CNRS, LIP6, F-75005 Paris, France \and \url{https://a-baudin.github.io/} }{alexis.baudin@lip6.fr}{}{} 

\author{Lionel Tabourier}{Sorbonne Université, CNRS, LIP6, F-75005 Paris, France \and \url{https://lioneltabourier.fr/} }{lionel.tabourier@lip6.fr}{}{}

\author{Clémence Magnien}{Sorbonne Université, CNRS, LIP6, F-75005 Paris, France \and \url{https://lip6.fr/Clemence.Magnien/} }{clemence.magnien@lip6.fr}{}{}

\authorrunning{A. Baudin, L. Tabourier and C. Magnien} 

\Copyright{Alexis Baudin, Lionel Tabourier and Clémence Magnien } 



\category{} 

\relatedversion{} 

\acknowledgements{This work is funded by the ANR (French National Agency of Research) through the ANR FiT LabCom. The authors thank the SocioPatterns\,\footnote{\url{http://www.sociopatterns.org}} and Icon\,\footnote{\url{https://icon.colorado.edu}} communities, for making their datasets available.}

\nolinenumbers 

\EventEditors{Alexander Artikis, Florian Bruse, and Luke Hunsberger}
\EventNoEds{3}
\EventLongTitle{30th International Symposium on Temporal Representation and Reasoning (TIME 2023)}
\EventShortTitle{TIME 2023}
\EventAcronym{TIME}
\EventYear{2023}
\EventDate{September 25--26, 2023}
\EventLocation{NCSR Demokritos, Athens, Greece}
\EventLogo{}
\SeriesVolume{278}
\ArticleNo{3}

\begin{document}

\maketitle

\begin{abstract}
  Community detection is a popular approach to understand the organization of interactions in static networks. 
For that purpose, the Clique Percolation Method (CPM), which involves the percolation of $k$-cliques, is a well-studied technique that offers several advantages.
Besides, studying interactions that occur over time is useful in various contexts, 
which can be modeled by the link stream formalism.
The Dynamic Clique Percolation Method (DCPM) has been proposed for extending CPM to temporal networks.

However, existing implementations are unable to handle massive datasets.
We present a novel algorithm that adapts CPM to link streams,
which has the advantage that it allows us to speed up the computation time with respect to the existing DCPM method.
We evaluate it experimentally on real datasets and show that it scales to massive link streams.
For example, it allows
to obtain a complete set of communities in under twenty-five minutes for a dataset with thirty
million links, what the state of the art fails to achieve even after a week of computation. 
We further show that our method provides
communities similar to DCPM, but slightly more aggregated. 
We exhibit the relevance of the obtained communities in real world cases, and 
show that they provide information on the importance of vertices in the link streams.

\end{abstract}


\section{Introduction}
\label{sec:introduction}

Detecting communities in complex networks has been a focus of interest since the early years of the field to the point that even the number of surveys on the topic is large.
While the first ones aimed at giving a general view of the landscape (\textit{e.g.},~\cite{fortunato2010community}), more recent ones tend to focus on a particular issue, for instance the underlying purpose of the community detection~\cite{schaub2017many} or a specific family of networks~\cite{magnani2021community}.

The Clique Percolation Method (CPM) proposed by Palla \textit{et al.}~\cite{palla2005uncovering} is a well studied technique to detect communities in a graph.
It is appreciated for the advantages that its definition confers: the communities are defined locally and in a deterministic way and there is no need to use heuristics or optimization functions that are hard to interpret.
Also, it allows the communities to overlap each other, by contrast with most other techniques which lead to a partition of the nodes.
It is a desirable property in general as the frontier between communities is often difficult to decide.
While the early implementations were computationally costly, it was later improved~\cite{kumpula2008sequential,reid2012percolation} and the most recent one scales up to graphs with hundreds of millions of nodes and edges~\cite{baudin2022clique}.

Considering temporal networks, where the structure evolves dynamically, 
a standard approach consists in examining them as a sequence of snapshots and
run graph community detection algorithms on each of them.
Then, it makes sense to match the communities obtained from a time step to the next to obtain consistent groups through time~\cite{aynaud2010static}.
Another approach following the same purpose is to design communities that ensure cohesive structure continuity over a time interval~\cite{tang2022reliable}.
This strategy has been investigated for various applications, including mobile communication networks~\cite{nguyen2011overlapping} or social networks analysis~\cite{rossetti2017tiles}.
However, the instability of many community detection methods makes this task hard to achieve properly~\cite{cazabet2017dynamic}.
In addition, some works stress the importance of achieving online community detection, in which case the communities are updated at each time step, by aggregating new information to the existing communities~\cite{cui2013online,pan2014online}.
If the method is fast enough, it is possible to achieve a streaming community analysis of the data.
However, this type of method often comes at the cost of losing part of the long term meaning that a community might have.

CPM can be implemented on graph snapshots and avoid instabilities from one step to the next due to its deterministic nature; it thus appears as a good candidate for the approach described above. 
It was in fact proposed quite early on, in~\cite{palla2007quantifying}; we refer to this approach as the {\em Dynamic Clique Percolation Method} (DCPM).
Recently, Boudebza \textit{et al.}~\cite{boudebza2018olcpm} introduced a faster algorithm to do this, called {\em Online Clique Percolation Method} (OCPM).
However, describing a temporal network as a sequence of snapshots has shortcomings.
Indeed, it misses the fact that the time step of analysis chosen is frequently arbitrary, while in general the data do not exhibit an obvious timescale of analysis.
This is why formalisms to describe temporal networks have been developed to circumvent these limitations~\cite{casteigts2012time,holme2012temporal,latapy2018stream}. 
Among those, the link stream formalism stresses the symmetric roles of structure and time in the representation of data, and aim at describing temporal networks at the intersection of graph theory and time series analysis~\cite{latapy2018stream}.
As the notion of clique has been recently extended to this formalism
and the search for cliques is implemented efficiently~\cite{baudin2023faster,viard2018enumerating},
it is now possible to investigate extensions of CPM to link streams.
This makes possible to design faster algorithms than the state-of-the-art DCPM implementations.
In this way, we obtain a community detection technique that can process data online and maintain relevant communities which naturally spread through time.

This is the main contribution of this paper: we propose an algorithm for this goal as well as an open source implementation that scales to large link streams\,\footnote{\url{https://gitlab.lip6.fr/baudin/lscpm}}.
In Section~\ref{sec:definitions}, we give the necessary background definitions and notations.
Then, we describe our method in Section~\ref{sec:algorithm} and derive an expression of its theoretical complexity; note that this method uses a novel $k$-clique enumeration algorithm in link streams.
Finally, we provide in Section~\ref{sec:experiments} an extensive experimental investigation which shows its efficiency on several real-world instances,
compares the obtained communities to the DCPM ones,
and illustrates its relevance to draw information about the data examined.

\section{Definitions and notations}
\label{sec:definitions}

First, we start with some basic reminders about (static) graphs. A graph is a pair $G = (V,E)$, where $V$ is a set of vertices, and $E$ is a set of edges of the form $\{u,v\}$, where $u,v \in V$ and $u \neq v$.
A $k$-clique in $G$ is a set of $k$ vertices that are all connected pairwise by an edge.
Finally, the definition of CPM communities is given by the fact that two $k$-cliques are adjacent when they have $k-1$ vertices in common. Then, a CPM community of $G$ is the set of vertices belonging to a maximal set of $k$-cliques that can be reached from one to another by a series of adjacent $k$-cliques (it forms a connected component of the graph whose vertices are the $k$-cliques of $G$ and edges are defined by the adjacency relation just explained).

\subsection{Cliques in link streams}

In this article, we work with the link stream model, which represents interactions over time.
Formally, a {\em link stream} is a triplet $L = (T,V,E)$ where $T$ is a time interval, $V$ a set of vertices and $E \subseteq T \times T \times V \times V$ a set of links $(b,e,u,v)$ such that $e \geq b$;
we call $e-b$ the duration of such a link.
Throughout the paper, we consider link streams with no self-loop, \ie for any link $(b,e,u,v) \in E$, then $u \neq v$. 
Moreover, links on the same vertices exist over disjoint time intervals, \ie if $(b,e,u,v), (b',e',u,v) \in E$, with $b \neq b'$ or $e \neq e'$, then $[b,e] \cap [b',e] = \emptyset$.

We use the definition of a clique in a link stream which follows the one in~\cite{viard2016computing}, with a minor difference to avoid cliques over time intervals of null length:
a {\em clique} is a pair $(C,[t_0,t_1])$, where $C \subseteq V$, $|C| \geq 2$ and $t_0,t_1 \in T$, $t_0 < t_1$, such that for all $u,v \in C, u \not=v$, there is a link $(b,e,u,v)$ in $E$ such that  $[t_0,t_1] \subseteq [b,e]$.
A {\em $k$-clique} is a clique containing $k$ vertices.
Notice that if $(C,[t_0,t_1])$ is a $k$-clique, then
$(C,[t'_0,t'_1])$ is also a $k$-clique for all $t'_0, t'_1$
such that $t_0 \le t'_0 < t'_1 \le t_1$. 
We are therefore interested in maximal $k$-cliques:

\begin{definition}[maximal $k$-clique]
  For $ k \in [\![ 2 , + \infty [\![ $, a \emph{maximal $k$-clique} is a clique $(C,[t_0,t_1])$ having $k$ vertices ($|C|=k$), and such that its time interval is maximal: there is no $t_0'<t_0$ nor $t_1'>t_1$ such that $(C,[t_0',t_1])$ or $(C,[t_0,t_1'])$ is a clique. 
\end{definition}
With this definition, we can introduce the notion of  $k$-clique adjacency, which will allow defining a generalization of CPM communities to link streams:
two maximal $k$-cliques $(C,[t_0,t_1])$ and $(C',[t_0',t_1'])$ are said to be \emph{adjacent} if they share $k-1$ vertices and overlap over a time interval with strictly positive length, \ie, $|C \cap C'| = k-1$ and $| [t_0,t_1] \cap [t_0',t_1'] | > 0$, where $|I|$ denotes the length of interval $I$.

\subsection{Communities in link streams}

In a dynamical context, it is natural to define a \emph{temporal community} as a set of {\em temporal vertices} of the form $(u,I)$, where $u$ is a vertex, and $I$ is a set of disjoint time intervals, which are the time intervals during which $u$ is present in the community.
Then, the notion of LSCPM community is similar to the one of CPM community in graphs, but with the notion of maximal $k$-clique adjacency adapted to link streams:

\begin{definition}[LSCPM community]
  A \emph{LSCPM community} is a temporal community whose temporal vertices belong to a maximal set of $k$-cliques that can be reached from one to another by a series of adjacent $k$-cliques.
\end{definition}

A few observations can be made about this definition.
First, as $k$ increases, the communities may only split and/or lose temporal vertices. 
In other words, if $k_1<k_2$,  each community computed with $k_2$ is included in a community computed with $k_1$.
This property is illustrated and further discussed in Section~\ref{subsec:kinfluence}.

Second, let us recall the definition of the dynamic CPM (DCPM) communities introduced in~\cite{palla2007quantifying}.
The idea is to compute the CPM communities at each snapshot.
Then, comparing the communities obtained for two consecutive snapshots allows finding whether each community evolves
by gaining vertices, losing vertices, dying (disappearing or merging with a larger one), or being born (appearing or being detached from a larger one).
Notice that DCPM communities can be considered as temporal communities: given two consecutive snapshots at $t_i$ and $t_{i+1}$, a vertex of a CPM community at $t_i$ belongs to the DCPM community on $[t_i,t_{i+1}[$. 
Most importantly, a LSCPM community is a union of DCPM communities.
Indeed, the temporal vertices of a CPM community in a snapshot are all included in a same LSCPM community; 
and a CPM community that gains or loses vertices from one snapshot to the next remain included in the same LSCPM community.
Note that two DCPM communities that are merged (resp. split) in the next snapshot belong to the same LSCPM community.

\begin{figure}[!hbt]
  \centering
  
  \subfloat[\centering Example of a link stream. \label{subfig:ls}]{{\includegraphics[width=0.49\linewidth]{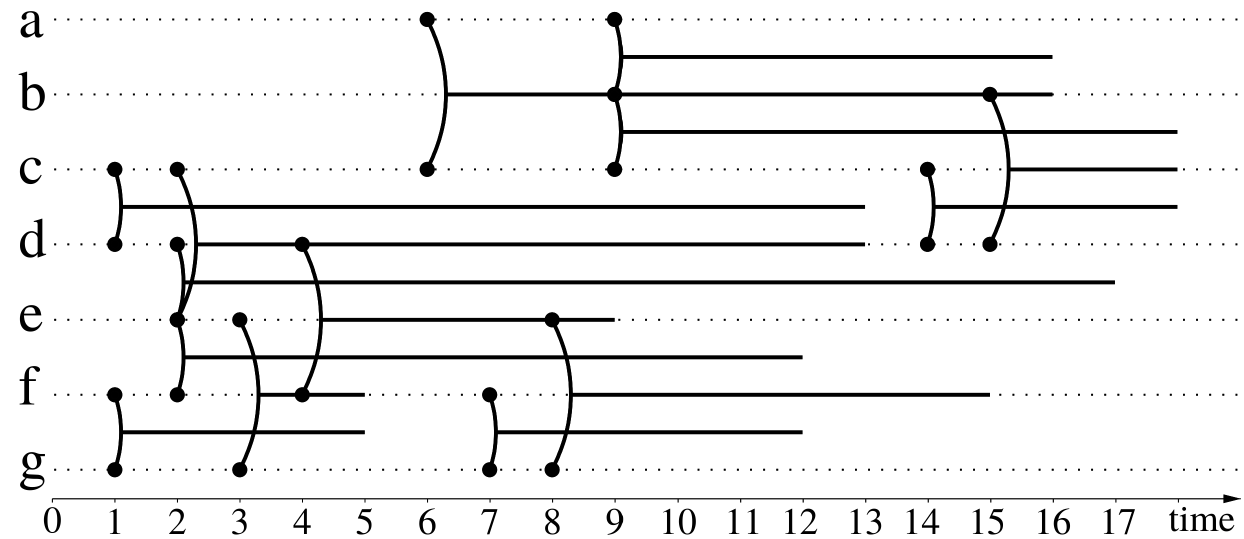} }}
  \hfill
  \subfloat[\centering Link stream of Figure~\ref{subfig:ls} and all its maximal 3-cliques in color. \label{subfig:lskcliques}]{{\includegraphics[width=0.49\linewidth]{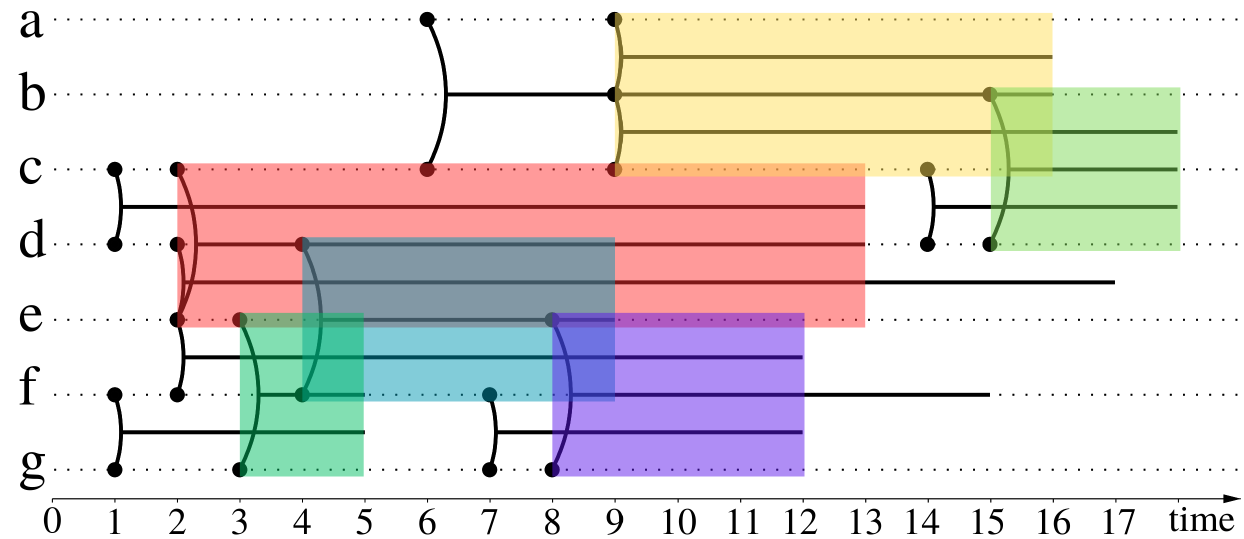} }}

  \subfloat[\centering Link stream of Figure~\ref{subfig:ls} and its two LSCPM communities in color. \label{subfig:lscpm}]{{\includegraphics[width=0.49\linewidth]{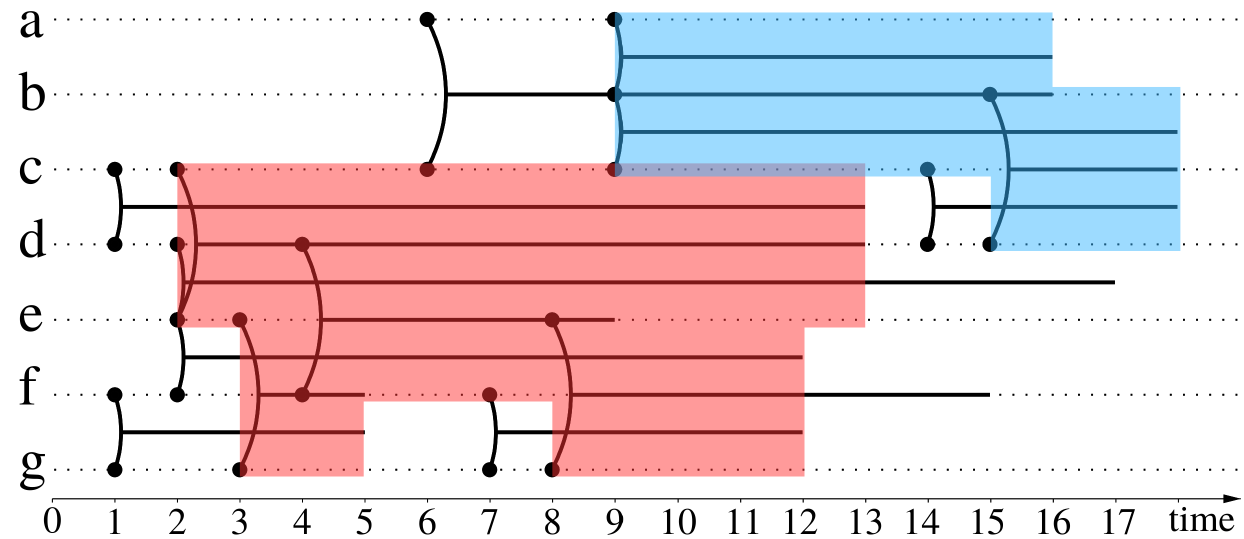} }}
  \hfill
  \subfloat[\centering Link stream of Figure~\ref{subfig:ls} and its four DCPM communities in color. \label{subfig:dcpm}]{{\includegraphics[width=0.49\linewidth]{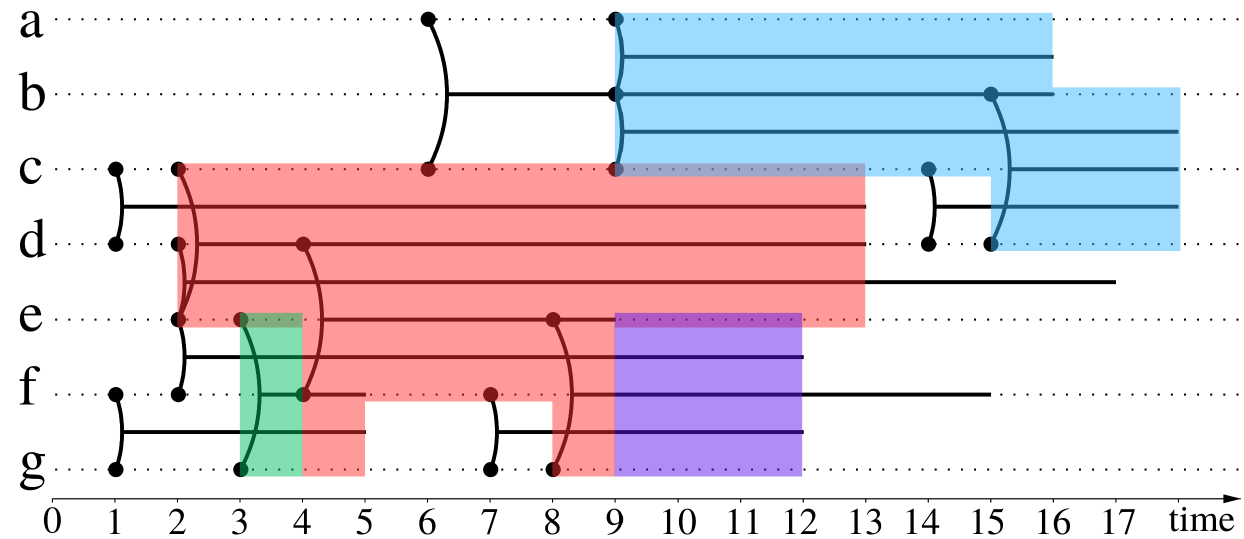} }}

  \caption{Example of a link stream with its maximal $k$-cliques for $k=3$ and the associated LSCPM communities and DCPM communities.} 
  \label{fig:kclique-ocpm-lscpm}
\end{figure}

To illustrate these definitions, we show in Figure~\ref{subfig:ls} an example of a link stream, with time on the X-axis and vertices on the Y-axis. 
Links are represented as connections between two vertices, and the horizontal line indicate their duration. 
For example, there is a link between vertices $c$ and $d$ over the time interval $[1,13]$. 
In Figure~\ref{subfig:lskcliques}, we represent its maximal $k$-cliques in color,
in Figure~\ref{subfig:lscpm} its LSCPM communities,
and in Figure~\ref{subfig:dcpm} its DCPM communities.
The background of each vertex is colored according to the time during which it belongs to its clique or community. 
Notice that the red LSCPM community is composed of three DCPM communities: the red one, but also the green one because $\{e,f,g\}$ is not adjacent to $\{c,d,e\}$ at time $t=3$; and the purple one because $\{e,f,g\}$ is no longer adjacent to $\{d,e,f\}$ for $t \geq 9$.

\section{Algorithms}
\label{sec:algorithm}

Our main idea to compute efficiently LSCPM communities in link streams
is to use techniques similar to those developed for enumerating maximal cliques in link
streams~\cite{baudin2023faster,viard2018enumerating}.
Indeed, it is possible to enumerate $k$-cliques in a link stream efficiently by going through each link only once and 
aggregate adjacent ones using a temporal Union-Find data structure.
In the DCPM case however, the comparison of communities at each time step to detect their evolution is more demanding.
We present an algorithm to efficiently enumerate $k$-cliques in Section~\ref{sec:kclique-listing}.
We then adapt the best existing method for computing CPM communities, 
and propose an efficient algorithm that, given as input the $k$-cliques of a link stream, 
computes its communities, in Section~\ref{sec:algo-lscpm}.

\subsection{$\boldsymbol{k}$-clique enumeration algorithm in link streams}
\label{sec:kclique-listing}

We take inspiration from our recent work~\cite{baudin2023faster} which enumerates the maximal cliques to design the algorithm for enumerating $k$-cliques in a link stream.
The key idea is to use graph $k$-clique enumeration, as very efficient algorithms have been designed for this task~\cite{danisch2018listing}.
Then, we compute the starting and ending time of each induced maximal $k$-clique in the link stream.

\begin{algorithm}[!hbtp]
  \DontPrintSemicolon 
  \KwIn{Link stream $L = (T,V,E)$; $k \in [\![ 3 , + \infty [\![$.}
  \KwOut{All $k$-cliques of $L$ without duplicates.}
  $G \gets$ empty graph \label{line:lskclique:emptygraph} \;
  $\E \gets$  empty associative array \tcp*{$\E$ associates ending times to edges}
  \For
  {$(b,e,u,v) \in E$ sorted by increasing $b$\label{line:lskclique:forE}} { 
    Add edge $\{u,v\}$ to $G$ \label{line:lskclique:addedge} \;
    $\E(u,v) \gets e$ \label{line:lskclique:Euv} \tcp*{Record the ending time of $\{u,v\}$}
    Remove from $G$ all edges $\{x,y\}$ with $\E(x,y) < b$   \label{line:lskclique:removeedge} \;
    $GCliques \gets $ all $k$-cliques of $G$ containing $u$ and $v$
    \label{line:lskclique:enumgraphcliques} \;
    \For {$C \in GCliques$\label{line:lskclique:forC}}{
      $end \gets \underset{x,y \in C}{\min} (\E(x,y))$ \label{line:lskclique:minEuv}\;
      \Output {$k$-clique $(C,[b,end])$} \label{line:lskclique:output} \;
    }
  }
  \caption{$k$-clique enumeration in link streams.}
  \label{algo:lskclique}
\end{algorithm}

Algorithm~\ref{algo:lskclique} lists all the maximal $k$-cliques.
It starts from an empty graph
(Line~\ref{line:lskclique:emptygraph}), and processes the link stream
in chronological order (Line~\ref{line:lskclique:forE}). 
For each link $(b,e,u,v)$, it updates the graph
(Lines~\ref{line:lskclique:addedge} to~\ref{line:lskclique:removeedge}). 
Then, it enumerates the maximal $k$-cliques containing $u$ and $v$
induced by the links seen up till then. 
They match the static $k$-cliques in $G$ that contain $u$ and $v$ (Line~\ref{line:lskclique:enumgraphcliques}).
Finally, each of these maximal $k$-clique starts at $b$, because the link between $u$ and
$v$ does not exist before $b$, and lasts as long as
all its links exist, so its ending time is the minimum of the ending
times of the edges composing it (Line~\ref{line:lskclique:minEuv}).

The complexity of Algorithm~\ref{algo:lskclique} is given by Theorem~\ref{thm:kclique-complexity}. 
Its proof is given in Appendix.

\begin{theoremApxrep}[$k$-clique enumeration time complexity]
  \label{thm:kclique-complexity}
  Algorithm~\ref{algo:lskclique} enumerates all $k$-cliques of the input link stream in time $\O{m \cdot k^3 \cdot \left(\frac{d}{2}\right)^{k-2} + m \cdot d^2 + m \cdot \log(m)}$,
  where $m$ is the number of links and $d$ the maximal degree of a node, that is its maximal number of neighbors at any given time.
\end{theoremApxrep}

\begin{proof}

  First, we show that the update of $G$ from Lines~\ref{line:lskclique:addedge}
  to~\ref{line:lskclique:removeedge} is done in $\O{m \cdot \log(m)}$ in the
  whole loop of Line~\ref{line:lskclique:forE}.
  $G$ is stored as an associative array, associating to each vertex the set of its neighbors,
  so the addition of an edge at Line~\ref{line:lskclique:addedge} is done in constant time. 
  Line~\ref{line:lskclique:Euv} is also done in constant time.
  Finding all edges that have an ending time smaller than $b$ (Line~\ref{line:lskclique:removeedge})
  can be done by 
  maintaining a sorted list of the end times of links in $G$,
  which can be done  in $\O{\log(m)}$ for each new link. 
  Then, the deletion itself  is done in constant time by going through this list.
  The global complexity is therefore in $\O{m \cdot \log(m) + m} = \O{m \cdot \log(m)}$.

  Consider an iteration of the loop starting at Line~\ref{line:lskclique:forE} and $(b,e,u,v)$ its associated link.
  We need to compute all $k$-cliques of $G$ containing $u$ and $v$. 
  This is equivalent to computing the ($k-2$)-cliques of the graph induced by the common neighbors of $u$ and $v$, that we note $G(N(u) \cap N(v))$.
  Computing this induced subgraph is done in $\O{d^2}$.
  Then, the overall complexity of these operations over all iterations is in $\O{m \cdot d^2}$.
  
  Enumerating the $(k-2)$-cliques of $G(N(u) \cap N(v))$ depends on the value of $k$. 
  If $k=3$ it consists in enumerating vertices, which is in $\O{d}$. 
  If $k=4$, it is enumerating the edges, in $\O{d^2}$. 
  If $k \geq 4$, then we use the $k$-clique enumeration algorithm in graphs described in~\cite{danisch2018listing}. 
  In that paper, Theorem~5.7 gives the complexity of enumeration in $\O{k \cdot m \cdot \left(\frac{d}{2}\right)^{k-2}}$.
  Thus, the $(k-2)$-clique enumeration is in $\O{(k-2) \cdot d^2 \cdot \left(\frac{d}{2}\right)^{k-4}}$ and the overall complexity of Line~\ref{line:lskclique:enumgraphcliques} is in $\O{k \cdot \left(\frac{d}{2}\right)^{k-2}}$, for any value of $k$. 
  Note that this value sets an upper bound on the number of $k$-cliques enumerated by the loop iteration. Each of these $k$-cliques is then processed by the loop at Line~\ref{line:lskclique:forC}, in $\O{k \cdot (k-1)} = \O{k^2}$. Then, the total complexity of these operations in the iteration is in $\O{k^3 \cdot \left(\frac{d}{2}\right)^{k-2}}$,  
  it is thus in $\O{m \cdot k^3 \cdot \left(\frac{d}{2}\right)^{k-2}}$ for the entire loop. 

  Combining the cost of the above operations finally gives the result.
  Note that this result can be slightly refined by keeping $k \cdot (k-1) \cdot (k-2)$ instead of $k^3$, which may have a significant impact, since $k$ values are usually small (typically $\leq 10$).
\end{proof}

The largest factor in this complexity is $m$.
$d$ can in theory be large with respect to $m$ but, since it is the maximal number of neighbors of a node {\em at any given time}, it is small in practice.
It is therefore the value of $k$ that
determines how efficient the enumeration can be, as the factors $k^3$
and $\left(\frac{d}{2}\right)^{k-2}$ show that this method remains
efficient only for small values of $k$.

\subsection{LSCPM: CPM algorithm in link streams}
\label{sec:algo-lscpm}

To the best of our knowledge, the most efficient algorithmic implementation of CPM in graphs is~\cite{baudin2022clique}.
In a few words, this algorithm stores each CPM community as the set of the $(k-1) $-cliques composing its $k$-cliques. 
It builds these sets on the fly, by processing each $k$-clique one by one, testing to which community each of its $(k-1)$-cliques currently belongs.
Then, it merges these communities or creates a new one if needed. 
For this purpose, the algorithm uses a Union-Find data structure, as it is efficient to do these operations. 
It is a forest of trees, where each node corresponds to a $(k-1)$-clique, and each tree corresponds to a CPM community, identified by its root.
It has three intrinsic functions: $\UF.\Find{id}$ that
returns the root of the tree containing the node $id$,
$\UF.\Union{p,q}$ that performs the union of two trees by connecting their roots and returns the root of the new tree,
and $\UF.\MakeSet{}$ which creates a new tree
on a new root $q$, and returns this root.

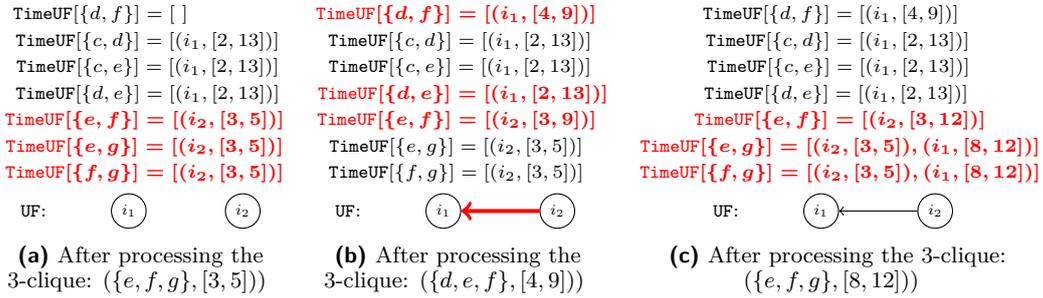
\begin{figure}[!hbt]
  \centering

  \scriptsize
  
  \subfloat[\centering {After processing the $3$-clique: $(\{e,f,g\},[3,5]))$} \label{subfig:clique1}]{{
      \begin{tikzpicture} 
        \tikzstyle{txt} = [scale=1]
        \tikzstyle{nod} = [circle, draw, fill=white, scale=0.8];
        \tikzstyle{gd} = [black,-];

        \node[txt] (a) at  (-0.62,2.6){$\timeUF[\{d,f\}] = [ \ ]$};
        \node[txt] (a) at  (0,2.25){$\timeUF[\{c,d\}] = [(i_1,[2,13])]$};
        \node[txt] (a) at  (0,1.9){$\timeUF[\{c,e\}] = [(i_1,[2,13])]$};
        \node[txt] (a) at  (0,1.55){$\timeUF[\{d,e\}] = [(i_1,[2,13])]$};
        \node[txt] (a) at  (-0.04,1.20){$\br{\timeUF[\{e,f\}] = [(i_2,[3,5])]}$};
        \node[txt] (a) at  (-0.04,0.85){$\br{\timeUF[\{e,g\}] = [(i_2,[3,5])]}$};
        \node[txt] (a) at  (-0.04,0.5){$\br{\timeUF[\{f,g\}] = [(i_2,[3,5])]}$};
        \node[txt] (a) at  (-1.5,0){\UF:};
        \node[nod] (1) at  (-0.25,0){$i_1$};
        \node[nod] (2) at  (1.25,0){$i_2$};
      \end{tikzpicture}
    }}
  \hfill
  \subfloat[\centering {After processing the $3$-clique: $(\{d,e,f\},[4,9]))$} \label{subfig:clique2}]{{
      \begin{tikzpicture} 
        \tikzstyle{txt} = [scale=1]
        \tikzstyle{nod} = [circle, draw, fill=white, scale=0.8];
        \tikzstyle{gd} = [black,-];
        
        \node[txt] (a) at  (-0.08,2.6){$\br{\timeUF[\{d,f\}] = [(i_1,[4,9])]}$};
        \node[txt] (a) at  (-0.03,2.25){$\timeUF[\{c,d\}] = [(i_1,[2,13])]$};
        \node[txt] (a) at  (-0.03,1.9){$\timeUF[\{c,e\}] = [(i_1,[2,13])]$};
        \node[txt] (a) at  (0.01,1.55){$\br{\timeUF[\{d,e\}] = [(i_1,[2,13])]}$};
        \node[txt] (a) at  (-0.07,1.2){$\br{\timeUF[\{e,f\}] = [(i_2,[3,9])]}$};
        \node[txt] (a) at  (-0.08,0.85){$\timeUF[\{e,g\}] = [(i_2,[3,5])]$};
        \node[txt] (a) at  (-0.08,0.5){$\timeUF[\{f,g\}] = [(i_2,[3,5])]$};
        \node[txt] (a) at  (-1.5,0){\UF:};
        \node[nod] (1) at  (-0.25,0){$i_1$};
        \node[nod] (2) at  (1.25,0){$i_2$};
        \draw [->,red,line width=1.5pt] (2) edge (1);
      \end{tikzpicture}
    }}
  \hfill
  \subfloat[\centering {After processing the $3$-clique: $(\{e,f,g\},[8,12]))$} \label{subfig:clique3}]{{
      \begin{tikzpicture}
        \tikzstyle{txt} = [scale=1]
        \tikzstyle{nod} = [circle, draw, fill=white, scale=0.8];
        \tikzstyle{gd} = [black,-];

        \node[txt] (a) at  (-0.12,2.6){$\timeUF[\{d,f\}] = [(i_1,[4,9])]$};
        \node[txt] (a) at  (-0.03,2.25){$\timeUF[\{c,d\}] = [(i_1,[2,13])]$};
        \node[txt] (a) at  (-0.02,1.9){$\timeUF[\{c,e\}] = [(i_1,[2,13])]$};
        \node[txt] (a) at  (-0.05,1.55){$\timeUF[\{d,e\}] = [(i_1,[2,13])]$};
        \node[txt] (a) at  (0,1.2){$\br{\timeUF[\{e,f\}] = [(i_2,[3,12])]}$};
        \node[txt] (a) at  (0,0.85){$\br{\timeUF[\{e,g\}] = [(i_2,[3,5]), (i_1,[8,12])]}$};
        \node[txt] (a) at  (0,0.5){$\br{\timeUF[\{f,g\}] = [(i_2,[3,5]), (i_1,[8,12])]}$};
        \node[txt] (a) at  (-1.5,0){\UF:};
        \node[nod] (1) at  (-0.25,0){$i_1$};
        \node[nod] (2) at  (1.25,0){$i_2$};
        \draw [->] (2) edge (1);
      \end{tikzpicture}
    }}
  \caption{Example of updates of \timeUF and \UF of Algorithm~\ref{algo:lscpm}, during the processing of the second, third and fourth $3$-cliques of the link stream of Figure~\ref{fig:kclique-ocpm-lscpm}. Note that all lists in \timeUF begin by a $(-1,-1,-1)$ triplet which we omit for readability.
    We show only the part of \timeUF relevant to the cliques under study. 
    At each time step, the $(k-1)$-cliques corresponding to the added clique are shown in red.
    In~\ref{subfig:clique1}, three $(k-1)$-cliques are added to the structure. 
    In~\ref{subfig:clique2}, communities of $i_1$ and $i_2$ are merged, as the $k$-clique contains one $(k-1)$-clique in $i_1$ and  another in $i_2$; also one $(k-1)$-clique is added: $\{d,f\}$, one is extended in time: $\{e,f\}$, and one remains unchanged because it is already present in the community over a longer time interval: $\{d,e\}$. 
    In~\ref{subfig:clique3}, one $(k-1)$-clique is extended in time: $\{e,f\}$, and as the other two, $\{e,g\}$ and $\{f,g\}$, were not in the community any more, they are re-added over the time interval of the $k$-clique $[8,12]$. 
    At the end of the process, the structure matches the information represented by the red LSCPM community of Figure~\ref{fig:kclique-ocpm-lscpm}. 
  }
  \label{fig:lscpm-example}
\end{figure}

We take inspiration from the algorithm above to extend the percolation of $k$-cliques to its definition in link streams.
The procedure is given in Algorithm~\ref{algo:lscpm} and follows a similar logic: each LSCPM community is stored as the set of the temporal $(k-1)$-cliques of its maximal $k$-cliques. 
A $(k-1)$-clique of a maximal $k$-clique $(C_k,[t_0,t_1])$ is of the form $(C_{k-1},[t_0,t_1])$, with $C_{k-1} \subseteq C_k$ containing $k-1$ vertices. 
These communities are constructed on the fly, by processing each maximal $k$-clique one by one, following the chronological order of their starting time, given by Algorithm~\ref{algo:lskclique}. 
For each maximal $k$-clique $(C_k,[t_0,t_1])$ (Line~\ref{line:lscpm:forCk}), Algorithm~\ref{algo:lscpm} checks the community to which each of its $(k-1)$-clique belongs on an interval that (strictly) intersects $[t_0,t_1]$ if it exists (Line~\ref{line:lscpm:adj}),
or creates a new one if needed (Line~\ref{line:lscpm:makeset}),
then merges them (Lines~\ref{line:lscpm:union} and~\ref{line:lscpm:append}).  
It also extends the membership duration of the $(k-1)$-cliques in case they did not belong to this community until $t_1$ (Line~\ref{line:lscpm:maxtime}).
To do so, in addition to the Union-Find structure \UF,
we use an associative array \timeUF, which associates each
$C_{k-1}$ to a list
of elements of the form $(id,[t_0,t_1])$, where $id$ is the identifier
of a Union-Find element and $[t_0,t_1]$ is the interval
during which $C_{k-1}$ belongs to the community of $id$.
In these lists, the intervals are disjoint, and the pairs are sorted in ascending chronological order.
Each list is initialized to [(-1,-1,-1)] (Lines~\ref{line:lscpm:forinit} and~\ref{line:lscpm:timeUFinit}), meaning that the corresponding $C_{k-1}$ has not yet been added to any community.
Figure~\ref{fig:lscpm-example} gives an example of the update of \timeUF and \UF structures, when applying Algorithm~\ref{algo:lscpm} to the link stream of Figure~\ref{fig:kclique-ocpm-lscpm}.

\begin{algorithm}[!hbtp]

  \DontPrintSemicolon 
  \KwIn{$(k-1)$-cliques then $k$-cliques of a link stream $L = (T,V,E)$; $k \in [\![ 3 , + \infty [\![$.}
  \KwOut{Union-Find structure representing all LSCPM communities of $L$.}
  \UF $\gets$ empty Union-Find data structure \label{line:lscpm:UF} \;
  \timeUF $\gets$ empty associative array \label{line:lscpm:nodesUF} \;
  \For {{\bf each} maximal $(k-1)$-clique $(C_{k-1},[t_0,t_1])$ of $L$\label{line:lscpm:forinit}}
  {
    $\timeUF[C_{k-1}] \gets [(-1,-1,-1)]$ \label{line:lscpm:timeUFinit}\;
  }
  \For {{\bf each} maximal $k$-clique $(C_k,[t_0,t_1])$ of $L$, sorted by increasing $t_0$\label{line:lscpm:forCk}} 
  {
    $p \gets -1$ \label{line:lscpm:p} \;
    \For {{\bf each} $u \in C_k$ \label{line:lscpm:forCk-1}}
    {
      $C_{k-1} \gets C_k \setminus \{u\}$ \; 
      $(id, [t'_0,t'_1]) \gets$ last element of $\timeUF[C_{k-1}]$ \label{line:lscpm:getid} \;
      \If (\tcp*[f]{is in the current community}) {$t_0 < t'_1$ \label{line:lscpm:adj}}
      {
        last element of $\timeUF[C_{k-1}]$ $\gets$ $(id,[t'_0,\max(t_1,t'_1)])$ \label{line:lscpm:maxtime}\; 
        $q \gets \UF.\Find(id)$ \;
        $p \gets \UF.\Union(p,q)$ \tcp*{merge with other $(k-1)$-cliques} \label{line:lscpm:union}
      }
      \Else (\tcp*[f]{not yet or no longer in the community})
      {
        \If {$p = -1$}
        {
          $p \gets \UF.\MakeSet{}$ \label{line:lscpm:makeset}\;
        }
        Append $(p,[t_0,t_1])$ to $\timeUF[C_{k-1}]$ \label{line:lscpm:append} \tcp*{add to community of $p$}
      }
    }
  }
  \caption{Clique Percolation Method in link streams (LSCPM).}
  \label{algo:lscpm}
\end{algorithm}

Finally, we need to transform the Union-Find structure into the adequate format to get the output as a temporal community.
This is done with a loop through the elements of \timeUF. Each one is of the form $C_{k-1} \rightarrow I$, where $C_{k-1}$ is a set of $k-1$ vertices, and $I$ is a list of pairs $(id,[t_0,t_1])$ corresponding to a Union-Find element and a time interval. 
Each Union-Find element belongs to a single set, which is one of the LSCPM communities and that we obtain with the \Find procedure.
We then add each vertex of $C_{k-1}$ to this community, on the time interval $[t_0,t_1]$.

The complexity of Algorithm~\ref{algo:lscpm} is given by Theorem~\ref{thm:lscpm-complexity} (demonstrated in Appendix). 
Note that Algorithm~\ref{algo:lscpm} takes as input the set of
$(k-1)$- and $k$-cliques of the link stream.
Therefore, the total complexity given in
Theorem~\ref{thm:lscpm-complexity}  takes into account 
the time needed to perform their enumeration
as well as the time needed to compute the communities.

\begin{samepage}
  \begin{theoremApxrep}[LSCPM time complexity]
    \label{thm:lscpm-complexity}
    The time complexity of Algorithm~\ref{algo:lscpm} is in $\O{(k + \alpha(n_k)) \cdot k  \cdot n_k + c(k)}$, where $\alpha$ is the inverse Ackermann function, $n_k$ the number of $k$-cliques of the link stream, and $c(k)$ the complexity of enumerating $k$-cliques, given in Theorem~\ref{thm:kclique-complexity}.
    It is thus in  $\O{\left(k + \alpha(n_k) \right)  \cdot m \cdot k^2 \cdot \left(\frac{d}{2}\right)^{k-2} + m \cdot d^2 + m \cdot \log(m)}$.
  \end{theoremApxrep}
\end{samepage}

\begin{proof}

  With a Union-Find data structure, it is known that the amortized cost of \Union and \Find functions is in $\O{\alpha(N)}$, if $N$ is the number of elements in the structure (see for example~\cite{tarjan1984worst}).
  Here, the Union-Find structure contains at most $1$ element per maximal $k$-clique,
  since there cannot be more than one call to \MakeSet (Line~\ref{line:lscpm:makeset}) in each iteration of the loop starting at Line~\ref{line:lscpm:forCk}.
  Indeed, if a \MakeSet is performed, then $p$ is no longer equal to $-1$
  and no other is performed until the end of this loop. 
  So the complexity of each call to \Union and \Find functions in the procedure is in $\O{\alpha(n_k)}$.

  Now, consider a maximal $k$-clique $(C_k,[t_0,t_1])$ corresponding to an iteration of the loop starting at Line~\ref{line:lscpm:forCk}. 
  Line~\ref{line:lscpm:forCk-1} performs one iteration per vertex of $C_k$, that is $k$ iterations.
  The operation at Line~\ref{line:lscpm:getid} to find the last element of  $\timeUF[C_{k-1}]$ is in $\O{k-1}$.
  During this loop, there is also at most one call to \Union and \Find, and other operations in constant time. 
  So, it runs in $\O{k+\alpha(n_k)}$. 
  Thus, in total, the loop starting at Line~\ref{line:lscpm:forCk} runs in $\O{n_k \cdot k \cdot (k+\alpha(n_k))}$.

  In addition, we have to take into account the complexity of the enumeration of $k$-cliques and $(k-1)$-cliques given as input to the algorithm.
  However, Theorem~\ref{thm:kclique-complexity} indicates that the complexity of enumerating $(k-1)$-cliques is included in the one of enumerating $k$-cliques, denoted $c(k)$.
  We thus obtain an overall complexity of Algorithm~\ref{algo:lscpm}  in $\O{(k + \alpha(n_k)) \cdot k  \cdot n_k + c(k)}$.  
  
  Finally, we saw in the proof of Theorem~\ref{thm:kclique-complexity} above that at each iteration of the loop at Line~\ref{line:lskclique:forE} of Algorithm~\ref{algo:lskclique}, the number of $k$-cliques listed is in  $\O{k \cdot \left( \frac{d}{2} \right)^{k-2}}$.
  Since there are $m$ iterations, we get that the number of $k$-cliques $n_k$ is in $\O{m \cdot k \cdot \left( \frac{d}{2} \right)^{k-2}}$.  
  Hence the second part of the theorem by combining the above bound on $n_k$ and Theorem~\ref{thm:kclique-complexity}. 
\end{proof}

\begin{toappendix}
  Note that the output of Algorithm~\ref{algo:lscpm} is the Union-Find structure and thus a post-processing is required to produce the actual communities.
  This post-processing is done by going through the set of elements $(C_{k-1},[t_0,t_1])$ in \timeUF, that are at most $k \cdot n_k$ (at most $k$ per maximal $k$-clique). Then, it performs a \Find operation on them and adds each of its $k-1$ associated vertices to its community during the corresponding time interval. 
  Adding a vertex to its community can be done in constant time if the nodes of the Union-Find are browsed in chronological order, which is possible by storing their creation time order at Line~\ref{line:lscpm:makeset}. So the complexity of the post-processing is also in $\O{k \cdot n_k \cdot (k + \alpha(n_k))}$.
\end{toappendix}

This complexity is expressed with the inverse Ackermann function $ \alpha $, which is known to grow extremely slowly, and can be considered as a constant at the scale of our data. 
Thus, we see from this theorem that the time complexity is close to
$\O{k^2 \cdot n_k + c(k)}$.
This shows that our algorithm is efficient in the way each $k$-clique is processed,
once the $k$-cliques have been computed.
Indeed, each $k$-clique contains $k$ $(k-1)$-cliques,
and therefore it is not possible to process them
in less than $k \cdot n_k$ operations, using an approach similar to ours.

The second part of the theorem is obtained by replacing $c(k)$ by the expression of Theorem~\ref{thm:kclique-complexity}
and $n_k$ by  a bound on its value.
If we do not take into account the factor $\alpha(n_k)$, the complexity is in  $\O{k^3 \cdot m \cdot
  \left(\frac{d}{2}\right)^{k-2} + m \cdot d^2 + m \cdot
  \log(m)}$. 
As noted above, this complexity depends almost linearly on $m$ and the 
factor $d$ is small in practice.
Nevertheless, the cubic factor in $k$ and the
  $\left(\frac{d}{2}\right)^{k-2}$
make it manageable only
for
  small values of $k$. We will see in Section~\ref{sec:experiments} that
  small values of $k$ allow for a fast building of the communities
    and are sufficient to observe interesting properties of the datasets.

In practice, Algorithm~\ref{algo:lscpm} needs to store in memory
  all the $(k-1)$-cliques in the link stream. It can be limiting, for
  example if the input dataset contains a very large clique. For instance, if 
  there is a clique of size $1000$, and that we are looking for $6$-clique
    communities, then there are more than $10^{15}$ $5$-cliques to store from this large clique. 
  Still, data from real-world interactions
  are known not to exhibit many large cliques, which makes the
  $k$-clique approach interesting for their study, and this memory
  feature
has not been prohibitive during our experiments.

\section{Experiments}
\label{sec:experiments}

For the experimental study, we implemented our algorithm in Python and the code is available online\,\footnote{\url{https://gitlab.lip6.fr/baudin/lscpm}}. 
Throughout this section, we set $k=3$ unless otherwise specified.
We will see that this value allows for a fast computation while being sufficient to provide interesting information on the datasets.
Also, we present in Section~\ref{subsec:kinfluence} the impact of increasing
$k$ on the community structures, which  induces smaller communities and therefore allows targeting their core,
with more or less strength depending on the value by which $k$ has
been increased.

\subsection{Datasets}
\label{subsec:dataset}

We run our experiments on real-world link streams of various sizes and types of interactions.
Even though many datasets
consist of links with duration,
in many cases these data are registered with regular discrete time intervals, 
because of the practical data acquisition protocol.
This is the case for instance of proximity between individuals data, usually captured using RFID tags. 
Therefore,
currently there is a larger range of publicly available instantaneous link streams with links of the form $(t,u,v)$, where $u$ and $v$ are vertices interacting at time $t$.
So, we transform these link streams by adding a duration $\Delta$ creating links of the form $(t,t+\Delta,u,v)$.
Note that the value of $\Delta$ will impact the number and extension of cliques in the link stream.
Practically, we choose uniform $\Delta$ values which are consistent with the typical time scales of the interactions considered for the datasets under study.
These values, while consistent, remain arbitrary,
and we use them to demonstrate the efficiency and relevance of our algorithm.
Users can adjust the values according to their requirements and to the nature of the studied datasets.

The datasets on which we performed the experiments are described in Table~\ref{tab:ls_data}. 
\households is a link stream representing contacts between members of five households in rural Kenya in 2012~\cite{data-contacts}; 
\highschool corresponds to contacts between students of five classes of a preparatory school in Marseilles (France) during one week in 2012~\cite{data-highschool2012} 
and \infectious consists of contacts between visitors to a museum in Dublin (Ireland) in 2009~\cite{data-infectious}. 
These three datasets represent contacts between individuals, for which we have chosen to take a link duration $\Delta=1$ hour.
The \foursquare dataset is extracted from the eponymous application, where users check-in in venues that they visit, located in New-York City in our case~\cite{data-check-in}.
It can thus be represented as a bipartite link stream between visitors and locations, where timestamps correspond to the check-in time.
In this link stream, we set $\Delta=6$ hours, then we project it on the set of locations:
if a user is connected to two locations over an overlapping time interval,
this will create a link between these locations during the overlap.
If the time interval of two links created in this way overlap, they are merged into a single link over the union of the initial time intervals.
Finally, we use the link stream \wikipedia, which represents links between Wikipedia pages, timestamped by the time of the link creation, over several years in the 2000s~\cite{data-wikipedia}.
We choose a $\Delta=1$ week duration, essentially to explore how our method scales to massive link streams.

\begin{table}[!hbt]
  \centering
  \begin{tabular}{|lc|ccccc|}
    \hline
    \textbf{Link stream} & $\boldsymbol{\Delta}$ & $\boldsymbol{m}$ & $\boldsymbol{n}$ & $\boldsymbol{d}$ & $\boldsymbol{D}$ & $\boldsymbol{r}$ \\
    \hline
    \households & 1 hour & 2,136 & 75 & 19 & 3 days & 1 hour \\
    \highschool & 1 hour & 5,528 & 180 & 18 & 8 days & 20s \\
    \infectious & 1 hour & 44,658 & 10,972 & 43 & 3 months & 5 min \\
    \foursquare & 6 hours & 268,472 & 33,153 & 81 & 10 months & 15 min \\
    \wikipedia & 1 week & 38,953,380 & 1,870,709 & 33,217 & 2.3 years & 20s \\
    \hline
  \end{tabular}
  \caption{Link stream datasets. $\boldsymbol{\Delta}$ is the link duration, $\boldsymbol{m}$ the number of links, $\boldsymbol{d}$ the maximal degree, $\boldsymbol{n}$ the number of nodes, $\boldsymbol{D}$ the total duration from the first to the last link and $\boldsymbol{r}$ the time resolution, that is the smallest duration between the beginning of two links.}
  \label{tab:ls_data}
\end{table}

The link stream parameter that affects most dramatically the running time of the community detection is its number of $k$-cliques $ n_k $, as the complexity depends  strongly on $k$ and $ n_k $ according to Theorem~\ref{thm:lscpm-complexity}. 
In Table~\ref{tab:nkcliques}, we report the number of $k$-cliques for each dataset, for $k$ ranging from 3 to 7. 
It allows anticipating the differences in computation time between the datasets, which are detailed in Section~\ref{subsec:expe-time}.
We notice that for large datasets, $ n_k $ increases with $k$, certainly because these datasets contain some large cliques.
Indeed,
within a clique containing $c$ vertices, there are $\binom{c}{k}$ cliques with $k$ vertices, and that quantity grows with $k$ (as long as $k \leq \frac{c}{2}$).
In particular, \foursquare is a projection of a bipartite network, and projections are known to contain many large cliques.

\begin{table}[!hbt]
  \centering
  \begin{tabular}{|l|ccccc|}
    \hline
    \textbf{Link stream} & $\boldsymbol{k=3}$ & $\boldsymbol{k=4}$ & $\boldsymbol{k=5}$ & $\boldsymbol{k=6}$ & $\boldsymbol{k=7}$ \\
    \hline
    \households & 3,951 & 4,721 & 3,929 & 2,324 & 987  \\
    \highschool & 2,468 & 583 & 97 & 11 & 1 \\
    \infectious & 79,836 & 128,157 & 202,181 & 274,181 & 300,850  \\
    \foursquare & 571,768 & 2,423,011 & 17,823,050 & 155,466,085 & 1,302,290,726 \\
    \wikipedia & 3,757,877 & 1,148,832 & 1,763,386 & 4,545,105 & 11,853,134 \\
    \hline
  \end{tabular}
  \caption{Number of $k$-cliques ($ n_k $) for each dataset and for $k$ from $3$ to $7$.}
  \label{tab:nkcliques}
\end{table}

\subsection{LSCPM: faster and scaling to massive real-world link streams}
\label{subsec:expe-time}

We now compare our algorithm to the DCPM one in terms of running time.
Note that the comparison focuses on the running time and not on the complexity as the complexity of the DCPM method or the existing OCPM implementation are not given by their authors.
In our implementation, the $k$-cliques are streamed to the standard input of the LSCPM algorithm, which  reads them as the enumeration proceeds. 
These two operations are done on two different threads; but for comparison purposes with the DCPM running time,
we measure its computation time as the sum of the time spent on each of these two threads. 
The DCPM runtime is provided by the best implementation available~\cite{boudebza2018olcpm}. 
We refer to this implementation as OCPM (standing for Online CPM).
We performed all the experiments on a Linux machine, equipped with two processors Intel Xeon Silver 4210R with twenty cores each, at 2.40Ghz, and with 252Gb of RAM.

\begin{figure}[!hbtp]
  \centering
  \includegraphics[width=0.3\linewidth]{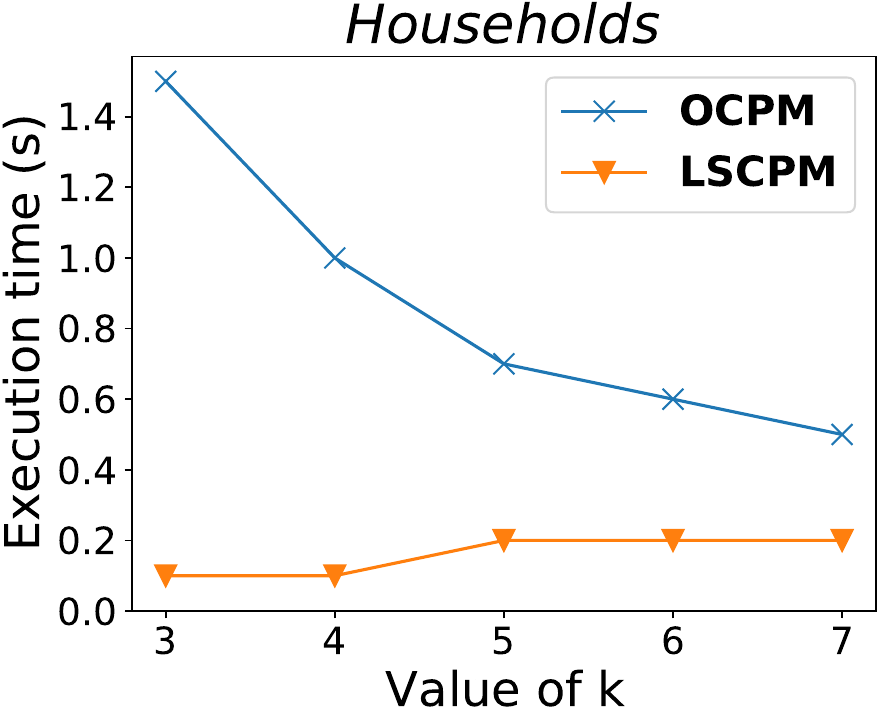}
  \quad
  \includegraphics[width=0.3\linewidth]{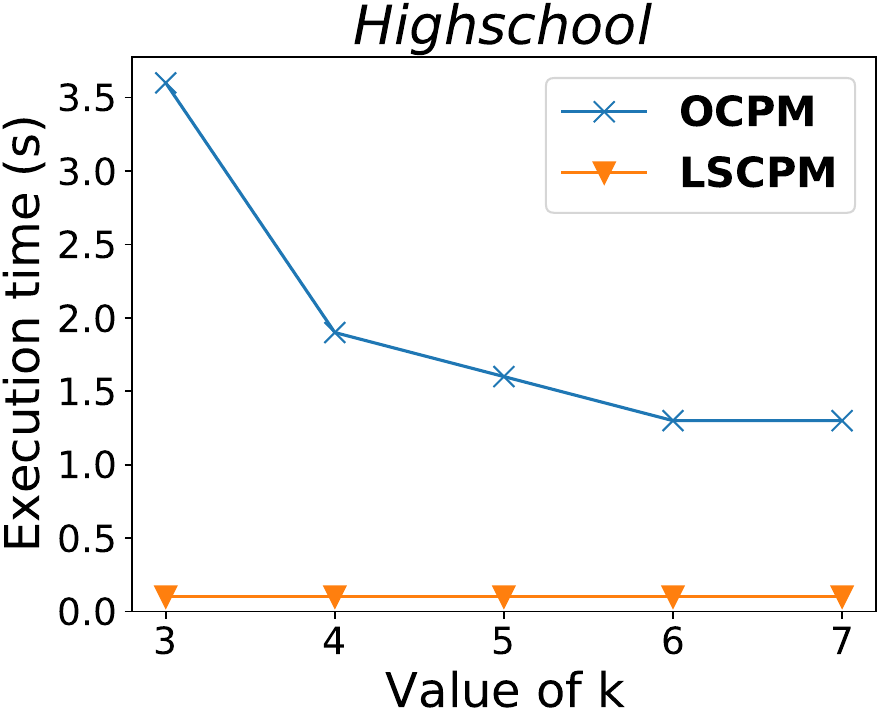}
  \quad
  \includegraphics[width=0.3\linewidth]{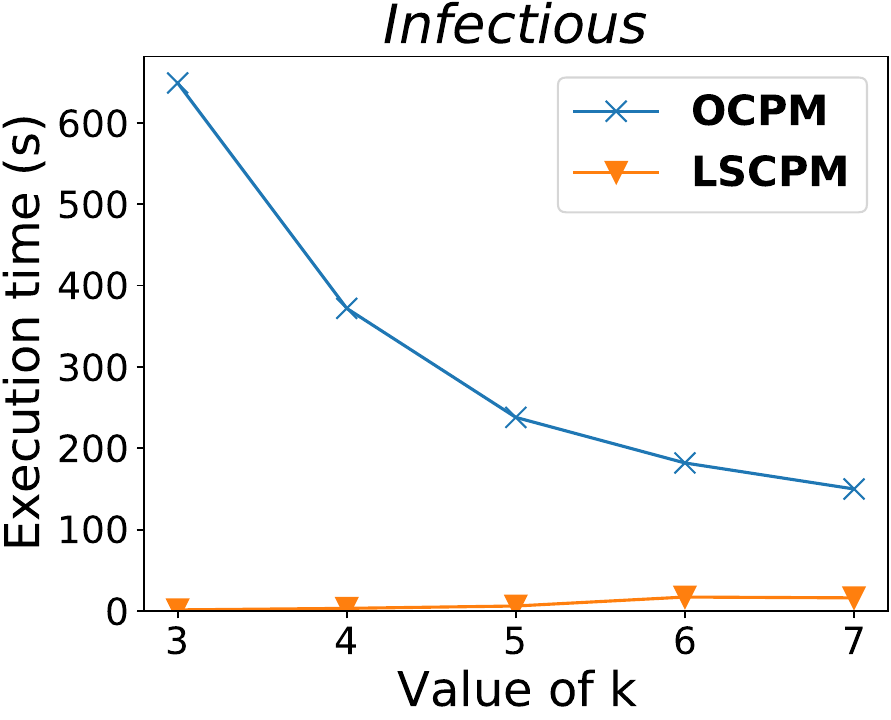}
  \quad
  \includegraphics[width=0.3\linewidth]{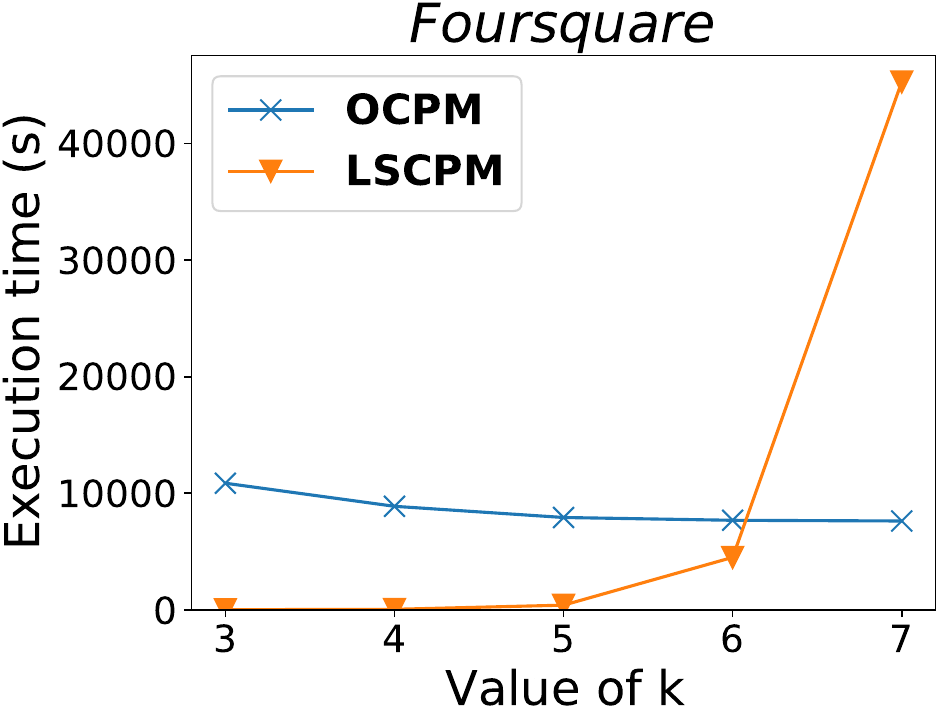}
  \quad
  \includegraphics[width=0.3\linewidth]{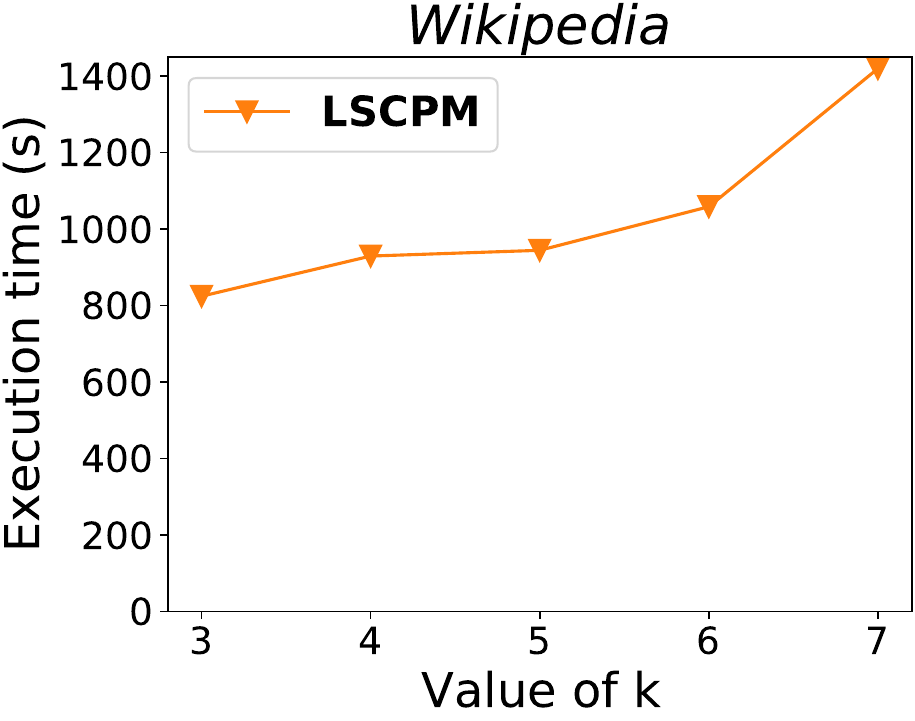}
  \caption{Line charts of the running times of OCPM and LSCPM for each dataset. Values are those in Table~\ref{tab:time}.}
  \label{fig:time-linechart}
\end{figure}

Table~\ref{tab:time} presents the computation times of communities with the OCPM implementation of DCPM and our LSCPM implementation, on all the datasets of Table~\ref{tab:ls_data}, for $k$ from 3 to 7.
These values are also plotted in Figure~\ref{fig:time-linechart} for readability purposes.
We observe that LSCPM is significantly faster, particularly on more massive datasets.
For example, with $k=3$ or $k=4$, it takes a few seconds with LSCPM to compute the \foursquare communities, while it takes a few hours with OCPM. 
Moreover, for the massive \wikipedia link stream, OCPM is unable to compute the set of communities in a week, while our algorithm provides the communities in less than 30 minutes for all $k$ values tested. 
Our algorithm thus allows to study a community structure in massive datasets for which the state of the art does not provide a result.

\begin{table}[!hbt]
  \centering
  \setlength{\tabcolsep}{2pt}
  \resizebox{\linewidth}{!}{
    \begin{tabular}{|l|cc|cc|cc|cc|cc|}
      \cline{2-11}
      \multicolumn{1}{c}{}&\multicolumn{2}{|c|}{$\boldsymbol{k=3}$}&\multicolumn{2}{c|}{$\boldsymbol{k=4}$}&\multicolumn{2}{c|}{$\boldsymbol{k=5}$}&\multicolumn{2}{c|}{$\boldsymbol{k=6}$}&\multicolumn{2}{c|}{$\boldsymbol{k=7}$}\\\hline
      \multicolumn{1}{|c|}{\textbf{Link stream}}&\textbf{OCPM}&\textbf{LSCPM}&\textbf{OCPM}&\textbf{LSCPM}&\textbf{OCPM}&\textbf{LSCPM}&\textbf{OCPM}&\textbf{LSCPM}&\textbf{OCPM}&\textbf{LSCPM}\\\cline{1-2}
      \hline
      \households&1.5s&0.1s&1.0s&0.1s&0.7s&0.2s&0.6s&0.2s&0.5s&0.2s\\
      \highschool&3.6s&0.1s&1.9s&0.1s&1.6s&0.1s&1.3s&0.1s&1.3s&0.1s\\
      \infectious&10min49s&1.4s&6min12s&3.3s&3min58s&6.2s&3min02s&17.2s&2min30s&16.2s\\
      \foursquare&3h01min&9.2s&2h28min&43s&2h12min&6min39s&2h08min&1h15mins&2h07min&12h35min\\
      \wikipedia&-&13min44s&-&15min29s&-&15min44s&-&17min38s&-&23min39s\\
      \hline
    \end{tabular}
  }
  \setlength{\tabcolsep}{6pt} 
  \caption{Time computation of communities in seconds with both OCPM and LSCPM, 
  for all our datasets, with $k$ varying from 3 to 7. The symbol ``-'' means that the computation time exceeds one week.}
  \label{tab:time}
\end{table}

Another point of interest is that the LSCPM computation time increases with $k$, while it decreases with OCPM.
This  comes from the fact that OCPM implementation obtains its results by aggregating the maximal cliques of size {\em at least} $k$,
while our method enumerates $k$-cliques.
With larger $k$, there are fewer maximal cliques to enumerate and process, hence the decreasing computation time.
By contrast, we have seen in Section~\ref{subsec:dataset} that $n_k$ typically increases with $k$ for larger datasets, which implies that the computation time of LSCPM increases too according to Theorem~\ref{thm:lscpm-complexity}.
Notice however that in spite of this, there is only one instance where OCPM is faster than LSCPM: \foursquare link stream with $k=7$, 
which as we have seen as a very large number of $k$-cliques.

\subsection{Comparison between LSCPM and DCPM communities}
\label{subsec:comp-lscpm-dcpm}

\begin{figure}[!hbt]
  \centering
  \includegraphics[width=0.3\linewidth]{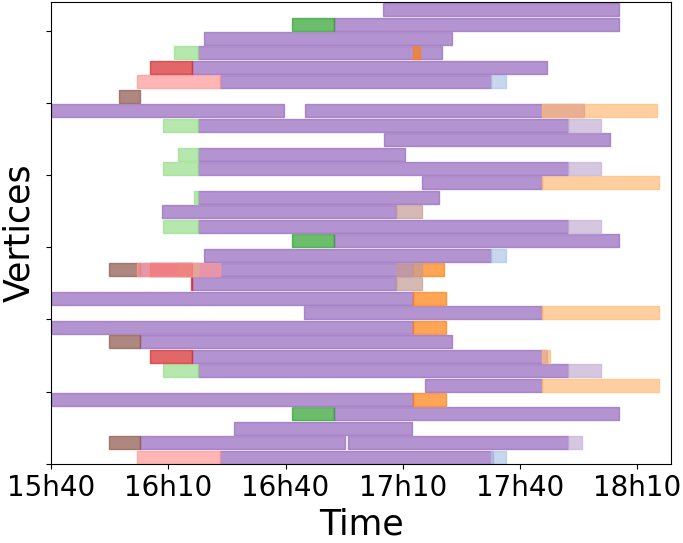}
  \hfill
  \includegraphics[width=0.33\linewidth]{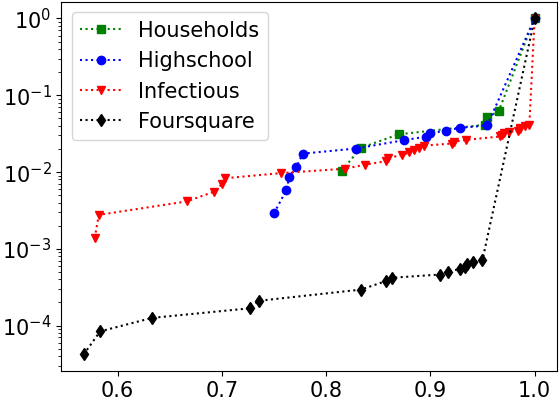}
  \hfill
  \includegraphics[width=0.315\linewidth]{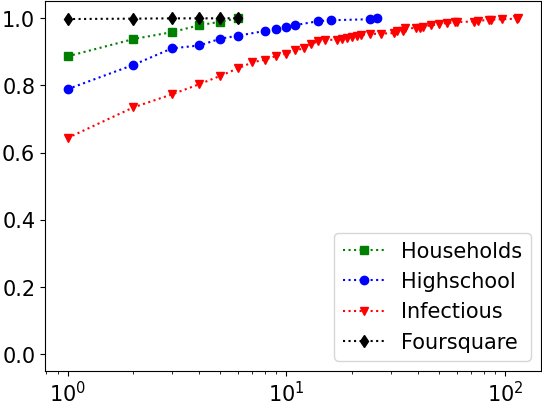}
  \caption{Composition of LSCPM communities in terms of DCPM communities, with $k=3$. 
  Left: a LSCPM community of \infectious, and the DCPM communities included in it (one color each). 
  Center: cumulative distribution of the relative size (in number of vertices) of the largest DCPM community to the corresponding LSCPM community. 
  Right: cumulative distribution of the number of DCPM community per LSCPM community.
  }
  \label{fig:inclusion-ocpm-lscpm}
\end{figure}

In what follows, we compare the communities obtained with our LSCPM algorithm to those obtained with DCPM, on the four datasets where the OCPM implementation provides a result.
Note that, up to our knowledge, there is no reference method to compare overlapping temporal communities. 
So we do not use tools such as NMI or Rand index which are designed for comparing partitions of vertices in a graph and thus would require some adaptation to the context considered in this paper.
We have seen in Section~\ref{sec:definitions} that each DCPM community is included in a LSCPM one and, conversely, that each LSCPM community can be seen as the union of DCPM communities. 
This property is illustrated in Figure~\ref{fig:inclusion-ocpm-lscpm}~(left), which gives an example of a LSCPM community from \infectious dataset, with $ k=3$ using the python package \texttt{tnetwork}\,\footnote{\url{https://tnetwork.readthedocs.io/}}.
Vertices are represented on the Y-axis, time on the X-axis, and each vertex belongs to the community over the period during which it is colored. 
Each of the DCPM communities included in this LSCPM community is represented in a different color. 
In what follows, we investigate to what extent DCPM communities are grouped into LSCPM communities.

To evaluate the similarity between a LSCPM community and the DCPM communities that it contains,
we compute the relative size (in number of vertices) of the largest DCPM community that each LSCPM community contains.
Figure~\ref{fig:inclusion-ocpm-lscpm} (center) reports the cumulative distribution of this value.
We see a clear peak at the end, which shows that for all datasets, $90\%$ of the LSCPM communities contain as many vertices as their largest DCPM community, which is the case of the example in Figure~\ref{fig:inclusion-ocpm-lscpm}~(left).
Only $1\%$ of LSCPM communities have its largest DCPM community with less than $70\%$ of the vertices, and none have its largest DCPM community with less than $50\%$.

We also observe  that there are only a few DCPM communities per LSCPM community.
It is illustrated in Figure~\ref{fig:inclusion-ocpm-lscpm} (right), which represents the cumulative distribution of the number of DCPM communities that each LSCPM community contains.
In all cases, more than $70\%$ of LSCPM communities contain only 1 or 2 DCPM communities, and almost never more than 10. 
However, there are some exceptions: in \highschool, $2.6\%$ of the communities contain between 10 and 26 DCPM communities, and in \infectious, $1.8\%$ of LSCPM communities contain between $50$ and $115$ DCPM communities.

Besides, we observe that there are fewer small LSCPM communities than DCPM ones:
considering communities with 5 or fewer vertices, we observe that
\households has $17\%$ more DCPM communities than LSCPM, 
\highschool has twice as many and \infectious has six times as many, 
(however, the sets of LSCPM and DCPM communities are very similar for \foursquare).
This indicates that small DCPM communities tend to be aggregated into larger LSCPM ones, as observed in 
the example of Figure~\ref{fig:inclusion-ocpm-lscpm}~(left).
These observations give the typical scheme of how LSCPM communities compare to the DCPM ones: a LSCPM community is in general composed of one large DCPM community which contains almost all the vertices, and possibly a few residual communities.
It also indicates that most of the meaning conveyed by the communities obtained in both cases should be closely related, but LSCPM method allows streamlining the community analysis by aggregating the smaller, less meaningful, communities into the larger ones.

\subsection{Insights on LSCPM communities}

To investigate the relevance of the temporal communities obtained, we have at our disposal metadata retrieved from the dataset repositories: families of \households, class of \highschool, and GPS coordinates of locations in \foursquare. 

In the case of the \households and \highschool datasets, which are based on person-to-person interactions,
we observe that the communities are homogeneous in terms of these categories, as could be expected.
Indeed, in the case of \households dataset, $95\%$ of communities are composed of members of only one family, and the remaining $5\%$ of two families. 
In \highschool, $70\%$ of communities are composed of only one class, $23\%$ of two classes, $6\%$ of three classes, and $1\%$ of four classes. 

\begin{figure}[!hbt]
  \centering
  \includegraphics[width=0.32\linewidth]{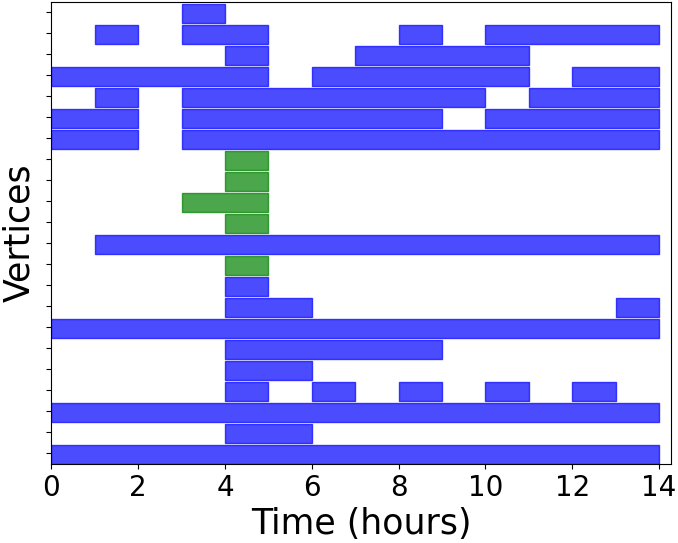}
  \qquad
  \includegraphics[width=0.32\linewidth]{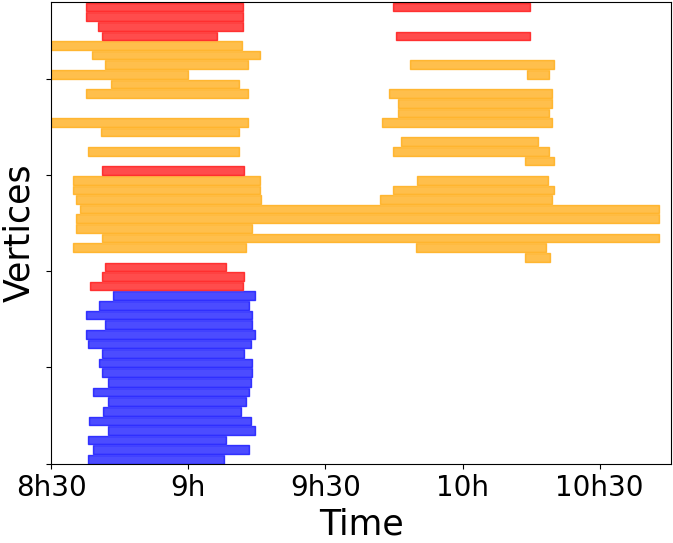}
  \qquad
  \includegraphics[width=0.24\linewidth]{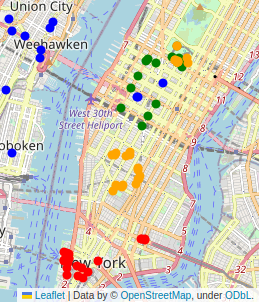}
  \caption{Examples of LSCPM communities using the metadata of the datasets.
  Left: a community of \households with vertices colored according to the family.
  Center: a community of \highschool with vertices colored according to the class.  
  Right: a map of Manhattan (NYC), where four \foursquare communities are represented in different colors.}
  \label{fig:metadata}
\end{figure}

Moreover, this metadata provides interesting insight on interactions at the individual level over time, pointing out who socializes with whom and when.
For instance, Figure~\ref{fig:metadata}~(left) is a community of \households, composed of 5 members of a family (in green) and 17 members of another (in blue).
It highlights the existence of an at most one hour gathering between all these people except one.
Similarly, Figure~\ref{fig:metadata}~(center) shows a community where we observe members of three different classes of \highschool, which are the three physics major classes of the preparatory school.
We distinguish three time periods: during the first one, students from the three classes are grouped together, which suggests a period when students can meet and mix.
Then, the community reduces to a few nodes of the orange class, later joined by an additional group of students, which might indicate the proximity of the students during the courses or working groups.

Concerning \foursquare, we can use the metadata to investigate the geographical distribution of locations which are visited by the same individuals within a $\Delta$ period.
Figure~\ref{fig:metadata}~(right) shows a map of a part of New-York City displaying a sample of four LSCPM communities. 
We observe that some of them are relatively clustered geographically, such as the green one, while others are more extended, which happens when one or several persons move from one part of the city to another within a $\Delta$ period. 
Thus here, $\Delta$ allows tuning the geographical extension of the communities, as lower $\Delta$ should correspond to smaller geographical extensions.

\begin{figure}[!hbtp]
  \centering
  \includegraphics[width=0.6\linewidth]{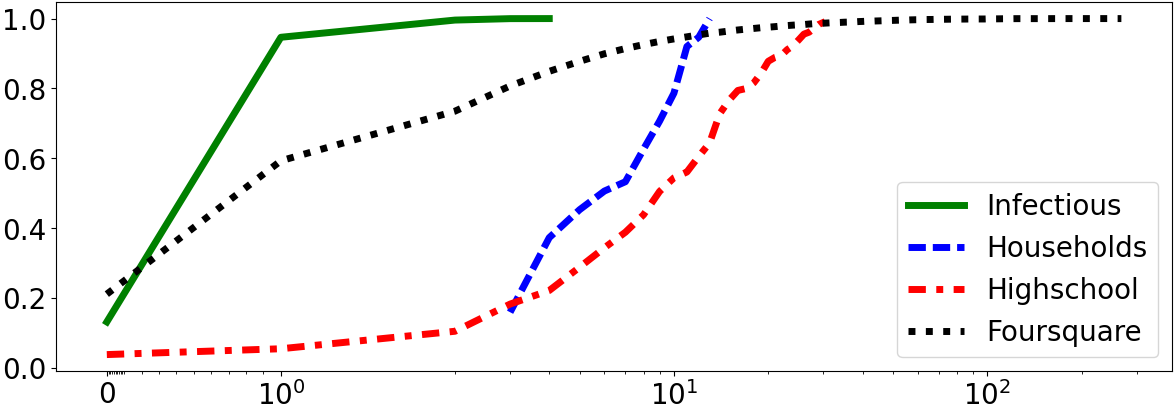}
  \caption{Number of communities to which each vertex belongs, as a cumulative distribution. 
  Note that the X-axis is in logarithmic scale, except between 0 and 1, in order to show vertices that belong to no community.}
  \label{fig:nodes}
\end{figure}

It may also be relevant to evaluate the involvement of vertices in the communities that they belong to.
Indeed, some vertices are active longer than others or belong to more communities, so it makes it possible to identify particularly important nodes in a group or nodes which act as bridges between groups.
Figure~\ref{fig:nodes} illustrates these two aspects: it is the cumulative distribution of the number of communities to which each vertex belongs.
Points on the far left correspond to vertices that belong to no community at all.
In this regard, the various datasets yield very different results.
For example, in \foursquare, around $20\%$ of the vertices do not belong to any community; it corresponds to locations where users rarely visit other places over the time period considered.
We observe this for some specific locations such as medical centers, offices, playgrounds...
By contrast, in \households, each vertex belongs to at least two communities, which is reasonable as it is a contact network between members of a same household, so no contact at all hardly seems conceivable.
In \highschool, we see that most nodes belong to many communities, which also makes sense, as students are grouped in classes, and each day makes new LSCPM communities.
Finally, vertices that belong to many communities are on the far right of the distribution.
It is striking on \foursquare, where almost $10\%$ of vertices belong to more than 10 communities, and a few to more than 100.
These can be described as central nodes of the link stream, which interact with many other nodes throughout the period of observation.
For example, the location of \foursquare that belongs to the most communities by far (1.5 times more than the second) is the famous Pennsylvania Station, which is the main intercity train station of New York City.
 
\subsection{Influence of $\boldsymbol{k}$ on the community structure}
\label{subsec:kinfluence}

The size $k$ of the cliques at the base of the LSCPM communities is the key parameter of the algorithm.
Here, we discuss the effect of increasing $k$ on the community structure, in order to give an intuition to the user as to the choice to make for this value.

\begin{figure}[!hbt]
  \centering
  \includegraphics[width=0.19\linewidth]{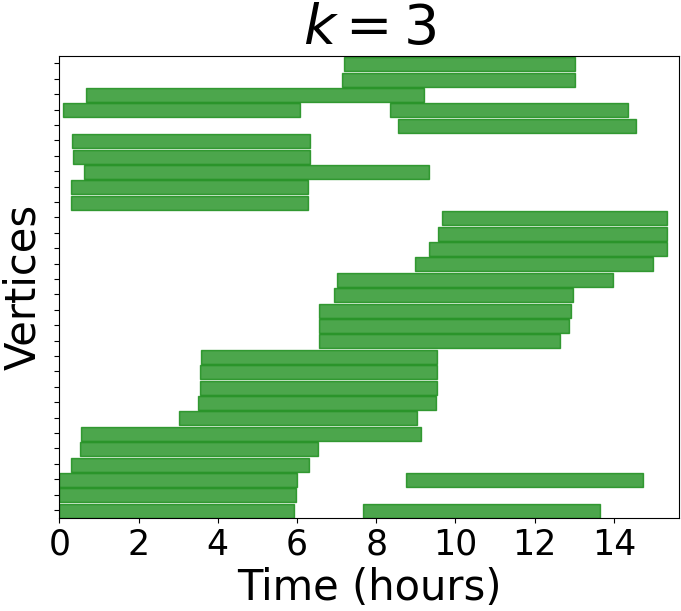}
  \hfill
  \includegraphics[width=0.19\linewidth]{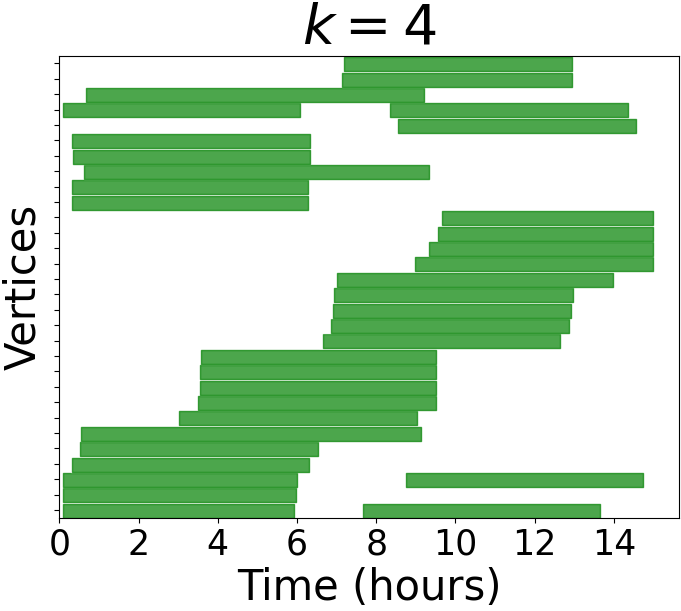}
  \hfill
  \includegraphics[width=0.19\linewidth]{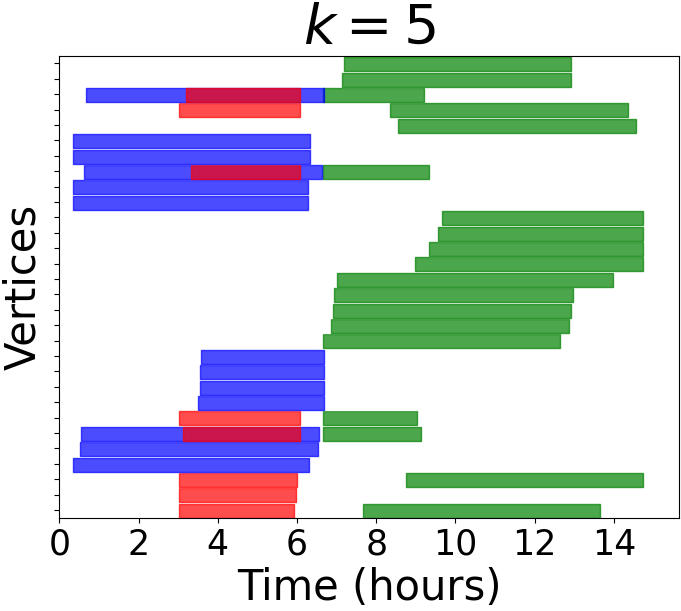}
  \hfill
  \includegraphics[width=0.19\linewidth]{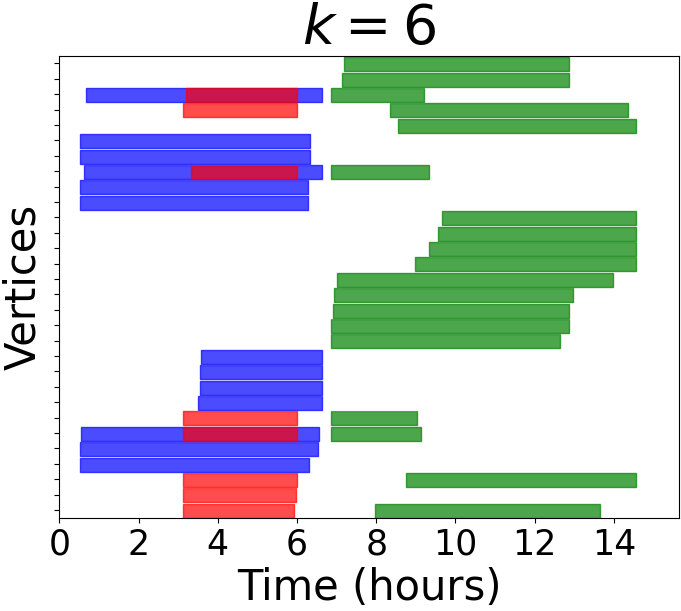}
  \hfill
  \includegraphics[width=0.19\linewidth]{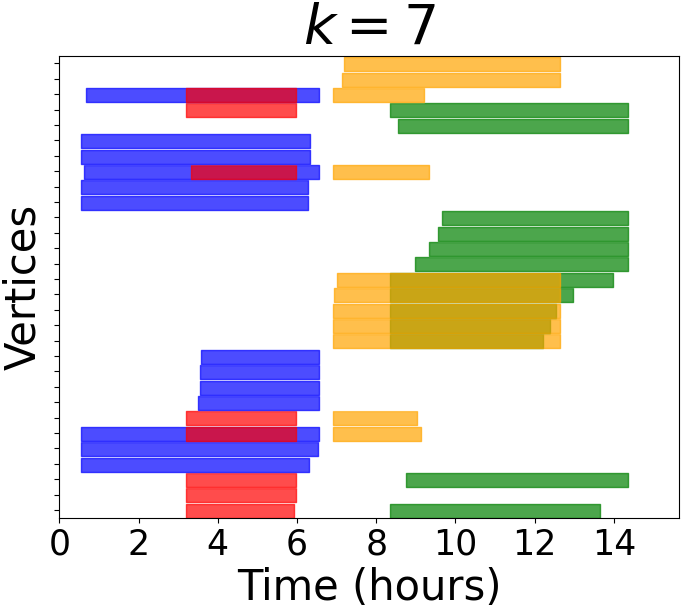}
  \caption{Split of a \foursquare community as $k$ increases from $k=3$ to $k=7$.}
  \label{fig:k}
\end{figure}

As we saw in Section~\ref{sec:definitions}, if $k_1 < k_2$, then each community computed with $k_2$ is included in a community computed with $k_1$. 
This means that increasing $k$ results in the communities splitting and/or loosing temporal vertices.
Figure~\ref{fig:k} gives an example of this phenomenon on a community of the \foursquare dataset: from left to right we see the community computed with $k=3$ and how it splits and looses nodes when increasing $k$ up to $7$.
We observe that the community remains almost identical for $k=4$, and that it splits into three communities for $k=5$, and one of these smaller communities split again for $k=7$.
We also observe that some nodes which belong to the community for $k=3$ at a given time do not belong to any of the resulting communities at this time for larger values of $k$.
Thus, increasing $k$ leads to more cohesive communities, but smaller in size and time extension.
This allows to change the granularity of the dynamical communities that can be obtained on a dataset by focusing on the ``core'' of the interactions. 

This allows to identify consistent sub-communities when metadata is available. 
For instance, the vertices of the \foursquare community of Figure~\ref{fig:k} correspond to 6 types of locations (out of the 261 available), all sport-related: Athletic-\&-Sport, Bike-Shop, Stadium, Sporting-Goods-Shop, Gym-/-Fitness-Center, Motorcycle-Shop (most other communities exhibit more diverse labels). 
We investigate if the splitting of communities when $k$ increases corresponds to more precise types of locations.
For $k=5$, the green community has the 6 type labels, but nodes of the blue and red ones only have 3 labels: Sporting-Goods-Shop, Gym-/-Fitness-Center and Bike-Shop. 
Also, for $k=7$, the green community splits into two parts, one of which focuses on two-wheeled sports: Motorcycle, Bike, Stadium, which highlights that the decomposition allows to derive the shared interests of the users.

\section{Conclusion and perspectives}
\label{sec:conclusion}

In this paper, we address the detection of dynamic communities in temporal networks, using the link stream formalism.
Using the literature of the field, we introduced the notion of maximal $k$-clique of a link stream, with an algorithm to enumerate them. 
This leads to a new adaptation of the Clique Percolation Method to dynamical networks, called LSCPM, which pursues the work initiated by~\cite{palla2007quantifying}.
We provided a theoretical analysis of the complexity of our algorithm, as well as an open source implementation in Python. 
Then, we experimented with the algorithm, comparing it to the state of the art, and showed that it allows to obtain possibly relevant information on real-world examples.

We believe that the community detection with LSCPM can scale to even more massive networks and larger values of $k$. 
For instance, as memory consumption can be a limiting factor on massive streams because of the clique storage, the work done in~\cite{baudin2022clique} to reduce the RAM cost of the CPM method on graphs could be adapted to link streams.
Also, it could be better in some cases to percolate maximal cliques instead of $k$-cliques, as done in OCPM~\cite{boudebza2018olcpm}, using efficient maximal clique enumeration methods in link streams such as~\cite{baudin2023faster}. 
In particular, we have seen that this may be an efficient alternative for larger $k$ values.

Moreover, it would be interesting to develop the analysis of the effect of link duration $ \Delta $ and its practical implications.
Indeed, when $\Delta$ increases, the $k$-cliques grow
longer, resulting in more aggregated LSCPM communities. 
This is the opposite effect of what happens when we increase $k$.
We believe that the experimental study can be extended by testing
the simultaneous tuning of these two parameters, to see how the
communities are structured, and if this allows targeting
relevant interaction cores in particular.

Finally, we believe that it is possible to adapt the algorithm for enumerating $k$-cliques in link streams, into a more general framework of temporal motif listing. 
Indeed, the call of the algorithm to enumerate graph cliques could be 
replaced in principle  by any other motif mining algorithm in a graph, resulting in a related temporal motif.
This paves the way to novel and efficient pattern mining algorithms in temporal networks.


\bibliography{biblio}

\counterwithin{figure}{section}

\end{document}